\documentclass[a4paper,fleqn,usenatbib]{mnras}

\usepackage{newtxtext,newtxmath}
\usepackage{lineno}
\usepackage{xcolor}

\usepackage[T1]{fontenc}
\usepackage{ae,aecompl}

\usepackage{graphicx}	
\usepackage{amsmath}	
\usepackage{multirow}
\usepackage{xcolor}
\usepackage{subfigure}
\let\oldAA\AA
\renewcommand{\AA}{\text{\normalfont\oldAA}}

\newcommand{\s}{S4\,0954+65}	
\newcommand{\txs}{TXS\,1515-273}
\newcommand{\rs}{RX\,J0812.0+0237}

\title[\s, \txs\ and \rs]{Optical spectral characterization of the gamma-ray blazars \s, \txs\ and \rs }

\author[Becerra Gonz\'alez]{J.~Becerra Gonz\'alez$^{1,2}$\thanks{jbecerra@iac.es}, J.~A.~Acosta--Pulido$^{1,2}$\thanks{jose.acosta@iac.es}, W.~Boschin$^{1,2,3}$, R.~Clavero$^{1,2}$, \newauthor J.~Otero--Santos$^{1,2}$, J.~A.~Carballo--Bello$^{4}$, L.~Dom\'inguez--Palmero$^{1,5}$\\
$^{1}$Instituto de Astrof\'isica de Canarias (IAC), E-38200 La Laguna, Tenerife, Spain\\
$^{2}$Universidad de La Laguna (ULL), Departamento de Astrof\'isica, E-38206 La Laguna, Tenerife, Spain\\
$^{3}$Fundaci\'on G. Galilei -- INAF (Telescopio Nazionale Galileo), Rambla J. A. Fern\'andez P\'erez 7, E-38712 Bre\~na Baja (La Palma), Spain\\
$^{4}$Instituto de Alta Investigaci\'on, Sede Esmeralda, Universidad de Tarapac\'a, Av. Luis Emilio Recabarren 2477, Iquique, Chile\\
$^{5}$Isaac Newton Group of Telescopes, Apartado 321, E-38700 Santa Cruz de La Palma, Canary Islands, Spain\\
}

\date{Accepted XXX. Received YYY; in original form ZZZ}

\pubyear{2020}
 
\begin{document}
\label{firstpage}
\pagerange{\pageref{firstpage}--\pageref{lastpage}}
\maketitle

\begin{abstract}
The study of gamma-ray blazars is usually hindered due to the lack of information on their redshifts and on their low energy photon fields. This information is key to understand the effect on the gamma-ray absorption due to either extragalactic background light and/or intrinsic absorption and emission processes. All this information has also an impact on the determination of the location of the emitting region within the relativistic jets. In this work a new optical spectroscopic characterization is presented for three gamma-ray blazars: \s, \txs\ and \rs. For all the three targets the redshift determination is successful, and for the first time in the case of \txs\ and \rs. Their classification as BL~Lac type is confirmed based on these new optical spectra. For \s\ (z=$0.3694\pm0.0011$) an estimation on the disk, broad line region and torus luminosities is performed based on the observed optical emission lines. The results from this study are compatible with the nature of \s\ as a transitional blazar. In the case of \txs\ ($z=0.1281\pm 0.0004$), although its optical spectrum is dominated by the continuum emission from the jet, applying the pPXF technique, the stellar population can be unveiled and is compatible with an old and metallic population. It is also the case of \rs\ ($z=0.1721\pm 0.0002$). Moreover, this work confirms that the optical spectrum from \rs\ is compatible with an extreme blazar classification.

\end{abstract}




\section{Introduction}\label{intro}

Due to geometrical effects, the emission from Active Galactic Nuclei (AGNs) whose relativistic jets point in the direction of the Earth (collectively called blazars) is boosted. Such relativistic jets emit continuum emission through the entire electromagnetic spectrum, from radio to gamma rays. Actually, the extragalactic gamma-ray sky is dominated by the emission from blazars. They constitute the most luminous gamma-ray sources with persistent emission, this fact together with the relativistic beaming effect make possible the detection of distant sources. In order to understand the particles acceleration in relativistic jets, and some of the most energetic processes in the Universe, the study of gamma-ray blazars is crucial. There are still many unknowns associated with such targets, for instance the processes responsible for the high energy emission from blazars, the jet's structure and the jet's composition among others \citep[see e.g. ][and references therein]{2019NewAR..8701541H}. 

The multiwavelength (MWL) spectral energy distribution (SED) from blazars typically displays a double peak structure, where the first peak is usually centered around the optical to X-ray bands, and the second one within the gamma-ray regime. The emission at low energies can be explained as synchrotron radiation from the relativistic electrons within the jet. While there is consensus on the origin of the low energy emission, different scenarios are considered for the origin of the high energy emission including leptonic and hadronic processes. Different subclasses of blazars can be considered depending on the position of the peaks, the flux ratio between the peak at high and low energies (usually called Compton Dominance -CD-) and their optical spectra. Blazars can be classified as BL~Lac objects (BL~Lacs) and Flat Spectrum Radio Quasars (FSRQs). BL~Lacs typically peak at higher energies than FSRQs, their CD is relatively small and display nearly featureless optical spectra. In BL~Lacs, the optical spectrum is mainly dominated by the continuum emission from the jet, showing only weak emission lines with equivalent widths at rest frame $|\rm{EW}_{\rm{rest}}| <5$\AA. On the other hand, FSRQs are typically characterized by a high CD, their SED peaks are located at low frequencies and their optical spectra show strong emission lines ($|\rm{EW}_{\rm{rest}}| >5$\AA). Therefore, from the perspective of the study of gamma-ray blazars, their optical spectrum provides crucial information on a) the redshift and b) the existence of external photon fields (e.g. the host galaxy or the Broad Line Region -BLR-) which can interact with the jet. The investigation of these two characteristics from three gamma-ray blazars \s, \txs\ and \rs\ is the focus of the present work.

Distant gamma-ray sources, as it is the case of blazars, suffer from absorption on their gamma-ray spectra as the radiation propagates through the Universe. The interaction between gamma rays and the diffuse Extragalactic Background Light (EBL) via pair production is the responsible mechanism for the absorption of the observed gamma-ray spectra \cite[see e.g.][]{2013APh....43..112D}. In order to infer the intrinsic gamma-ray spectra emitted by distant sources the knowledge of the redshift is essential, as the probability of interaction increases with distance. Moreover, such interaction is energy dependent, producing not only absorption but also spectral distortion. Unfortunately, the determination of the blazar's redshifts is not an easy task, due to the fact that the continuum emission from their powerful jets outshines the spectral features. For this reason, the host galaxy and/or different parts of the AGN cannot be easily observed from blazars. In this work we study the optical spectra from three unknown redshift gamma-ray blazars, providing important information in order to infer both the intrinsic gamma-ray spectra emitted by the target and the EBL imprint (which carries cosmological information).

Moreover, the gamma-ray spectra from blazars can suffer from intrinsic absorption due to the presence of low energy photon fields, as it is the case for the optical emission from the BLR for instance. The characterization of the optical photon field is therefore crucial not only to characterize the possible gamma-ray absorption via pair production, but also to evaluate the possible contribution to the particles acceleration up to gamma rays. Within the leptonic framework, gamma rays are produced via inverse Compton scattering of relativistic electrons with low energy photons e.g. optical photons. Therefore, the characterization of the optical spectrum from blazars allows us to estimate the optical external photon fields to the jet in case spectral features are found.

\s\ was detected in gamma rays by EGRET \citep{1995ApJ...445..189M} and later on with {\it Fermi}-LAT in the high energy range (HE, $E\geq100$ MeV) \citep[see][and references therein]{2016PASJ...68...51T}. During a very strong flare detected in almost all wavelengths in 2015, the source was also detected in the very-high-energy (VHE, $E\geq100$ GeV) band with the MAGIC telescopes \citep{2018A&A...617A..30M}. The redshift and classification of this source are still under debate. The first tentative redshift estimation was $z=0.368\pm0.004$ by \cite{1986AJ.....91..494L}, based on the preliminary identification of the [O~III] emission lines and Mg b absorption feature. Later on, \cite{1996ApJS..107..541L} firmly established the redshift as z=0.3668 inferred from the identification of [O~II], [O~III] and H$\alpha$ emission lines. Meanwhile, \cite{1993A&AS...98..393S} also confirmed the redshift but based on a different set of spectral features, Ca H\&K absorption doublet and a weak [O~II] emission feature. Due to the fact that both redshift estimates are derived from a different set of spectral features, new observations were carried out using the 10.4\,m Gran Telescopio de Canarias (GTC) by \cite{2015AJ....150..181L} aiming to confirm it. This new high signal-to-noise spectral observations taken in 2015 resulted in a featureless spectrum, and a lower limit of $z\geq0.45$ was derived. Therefore, these contradictory results keep still the redshift of \s\ under debate.

In addition to the redshift, the classification of \s\ is also controversial. Different authors have classified it as both BL~Lac and FSRQ blazar type, even it has been classified as a possible transitional blazar between these two categories \cite[see][and references therein]{2020ApJS..247...33A}. On the one hand, the target shows weak emission lines or lack of emission lines as later reported. On the other hand, the MWL SED displays a high CD and actually, the standard Synchrotron-Self Compton \citep[SSC,][]{1992ApJ...397L...5M} models which typically describe BL~Lac objects do not provide a successful approach. Instead, an external field is needed in addition to the SSC in order to explain the SED which shows a SED distribution more similar to FSRQs than BL~Lacs \citep{2016PASJ...68...51T, 2018A&A...617A..30M}. New optical spectroscopic observations are used in this work to investigate the properties of the source.

\txs\ is a gamma-ray blazar detected in the HE band by {\it Fermi}-LAT \citep[see e.g.][]{2020ApJS..247...33A}, classified as a BL~Lac. During a flaring state in February 2019, the source was detected in the VHE band with the MAGIC telescopes \citep{2019ATel12538....1M}. The only indication on its redshift comes from the photometric upper limit $z<1.1$ from \cite{2018ApJ...859...80K}. Optical spectroscopic observations on this object have also been carried out recently by \cite{2020arXiv201205176G}, obtaining compatible results with those presented here.

\rs\ is classified as a BL~Lac gamma-ray blazar in the 4FGL catalog \citep{2020ApJS..247...33A}. The source is also a VHE candidate contained in the 2WHSP catalog \citep[2WHSP\,J081201.7+023732,][]{2017A&A...598A..17C}. Although the synchrotron peak reported in the 2WHSP catalog is only $10^{11.6}$\,Hz based mainly on the lack of X-ray observations, its MWL SED shows a clear hint of the host galaxy. This fact points to the possibility that \rs\ is a candidate for an extreme blazar. Such extreme blazars (usually called EHBL), are blazars whose SED peaks are located at higher frequencies/energies in comparison with the typical SEDs displayed by blazars \cite[see e.g.][]{2020NatAs...4..124B, 2020MNRAS.494.6036B}. EHBLs are very interesting targets to be explored in the VHE regime as their emission budget is dominated by the VHE band. Moreover, due to the fact that their emission peaks are located at higher energies than usual, the host galaxy is unveiled in the optical range. This fact allows us to derive their redshift from the spectral optical features displayed by their host galaxy and characterize its optical emission. 

This paper is structured in five sections. In Sec.~\ref{sec:observations} the observations and data reduction procedures are detailed for the different data samples. The results for each individual target under study in this work are presented in Sec.~\ref{sec:results}. The conclusions derived from this work are summarised in Sec.~\ref{sec:conclusions}. Finally, some technical considerations on the significance of the detection of the spectral features are discussed in Appendix~\ref{appendix}.

\section{Observations and data reduction}
\label{sec:observations}

The optical observations presented in this work were carried out with three different telescopes at Roque de los Muchachos observatory located in the Canary island of La Palma. \s\ was observed during several epochs at different states of the source using the 3.58\,m Telescopio Nazionale Galileo (TNG). For \txs\ and \rs\ a single observation for each target was performed using the 2.5\,m Nordic Optical Telescope (NOT) and the 4.2\,m William Herschel Telescope (WHT), respectively. Standard calibration observations for bias, flat field and arc lamps were also taken during the same night for each of the target observations.

The data are reduced\footnote{IRAF standard packages are used for the different reduction tasks through this work.} following the standard procedures for bias subtraction and flat-field correction. The sky lines are subtracted averaging the sky spectrum as observed adjacent to the target. The spectra are flux calibrated using spectrophotometric standards taken after the target observations using the same configuration. Afterwards, the target spectra are corrected for telluric absorption using the task {\it telluric} available within IRAF packages. The telluric correction is derived from the spectrophometric standard spectrum,  after normalization of the continuum by a smooth function (polynomial). The stellar atmosphere absorption features are masked. We check for artifacts introduced by the telluric correction by comparison with the atmospheric transmission curves provided by the ESO SkyCalc tool \citep{Noll12}. 
Finally, the spectra are corrected using the  IRAF task {\it deredden} for interstellar reddening with the $A_V$ values provided by the NED database \citep{Schlafly11} and the extinction curve by \cite{CCM89}.

The details on the observations and analyses for each target are provided below. In particular, the observation setup for each target and the telescope used in this work are given. 

\subsection{\s}

The observations of \s\ were carried out using the DOLORES (Device Optimized for the LOw RESolution) spectrograph at TNG in long-slit configuration. Six observations were carried out using the LRB grism, the details of each observation can be found in Table~\ref{tab:obs_0954}. The He and Hg+Ne calibration lamps are used to calibrate the spectra in wavelength using the same instrumental setup as for the observation of the target and the calibration star. 

The flux calibration is performed by applying a spectral response function determined after combining the observations of the spectrophotometric standard stars taken on different nights (see Table \ref{tab:obs_0954}). This procedure is preferred to avoid introducing systematic differences coming for the use of different spectrophotometric standards. Each spectrum is previously corrected for atmospheric extinction using the extinction curve available for the observatory\footnote{\url{http://www.ing.iac.es/Astronomy/observing/manuals/ps/tech_notes/tn031.pdf}}. In order to confirm the absolute flux scale, each spectrum is rescaled with the $R$-band photometry obtained from the light-curve presented in \citet{2018A&A...617A..30M} for all dates, except for 2015 Dec 12 which is taken from Las Cumbres Observatory Global Telescope Network archive \citep{2013PASP..125.1031B}. The telluric corrections are performed using the spectrophotometric stars as described in Sec.~\ref{sec:observations}, except for the spectrum taken on 2015 April 23 for which a model provided by the ESO SkyCalc tool is used. This is due to the fact that the calibration star observed during that particular night (BD+75 325) displays intrinsic spectral features close to telluric absorption bands, complicating the correction.
Finally, the spectra are corrected for interstellar reddening using a value $A_V = 0.328$.

\begin{table*}
   \centering
   \begin{tabular}{c c c c c c c c c} 
   Date & MJD & Exposure & Slit & Airmass & Star & Seeing (star) & R@5500 \AA & S/N \\
        \hline
        \hline
    2015-02-24 & 57077.02 & 3x500\,s  & 1.0" & 1.25 & Hiltner 600 & 2.3"-2.5 (2.7") & 550 & 84 \\
    2015-04-23 & 57135.88 & 1x1000\,s & 1.0" & 1.25 & BD+75 325    & 0.8" (0.9") & 550 & 86 \\
    2015-05-19 & 57161.89 & 1x1800\,s & 1.5" & 1.33 & Feige 66    & 0.7" (1.4") & 367 & 105 \\
    2015-05-21 & 57163.90 & 3x1200\,s & 2.0" & 1.35 & Feige 34    & 0.6" (0.6") & 275 & 160 \\
    2015-06-27 & 57200.89 & 1x1800\,s & 1.5" & 1.75 & Feige 66    & 1.9" (1.7") & 367 & 113 \\
    2015-12-12 & 57368.13 & 1x900\,s  & 1.5" & 1.40 & Hiltner 600 & 0.9" (0.8") & 367 & 128 \\
        \hline
        \hline
   \end{tabular}
   \caption{Information on the long-slit spectroscopic observations of \s\ carried out at TNG with DOLORES using the LRB grism. The MJD date and the airmass are given for the start of the observation. The column named star refers to the flux calibration star observed for each target observation. The seeing value is given for the target observation, followed by the one for calibration star in brackets. The resolution reported in this table refers to the value obtained from the slit width. The S/N of the spectrum calculated in the range 4100-4600 \AA\ can be found at the last column. The best S/N is achieved on 2015 May 21, due to the fact that it was taken with the longest exposure and the best seeing conditions.}
   \label{tab:obs_0954}
\end{table*}

\subsection{\txs}
The spectrum of \txs\ was observed with the Alhambra Faint Object Spectrograph and Camera (ALFOSC) at NOT. The observations were carried out in long-slit configuration using the grism \#4 and 1" slit width yielding a resolution R=357 at 5500\.\AA. The observation was performed on 2019 May 7 (starting time MJD 58610.00) for a total of 1 h divided in 3 exposures of 1200\,s each. During the observation the seeing was reasonably stable from 0.9" to 1.0". Due to visibility constraints from the northern hemisphere, the target was observed at a high airmass of 2.02 at the beginning of the fist exposure up to 1.86 at the beginning of the last exposure. Right after the observation of the target, an exposure of the calibration arc of He+Ne was observed using the same instrumental configuration. The flux calibration is performed with the calibration star HD\,93521 \citep{1990AJ.....99.1621O} observed with the same instrumental setup as for the target. The telluric absorption features are removed using the spectrum of HD\,93521 as a template. Finally, the combined spectrum is corrected for interstellar reddening using $A_V = 0.646$.

The absolute flux level is checked using photometry obtained in $R$-band with the Kungliga Vetenskapsakademien (KVA) telescope. The data are analyzed using the differential photometry method described by \citet{2018A&A...620A.185N}. The KVA photometric observation was performed on 2019 May 6 (MJD 58609.95), $\sim$\,1\,h before of our spectroscopic observations with NOT, yielding a $R$-band magnitude $\simeq15.9$ (KVA team private communication). After the extinction correction, our flux calibrated spectrum agrees with the photometric measurement within a 5\% uncertainty.

\subsection{\rs}

The optical spectrum of \rs\ was obtained using ACAM (Auxiliary-port CAMera) at WHT on 2017 October 20 (starting time MJD 58046.24). The target was observed for a total of 300\,s, 3 exposures of 100\,s each. The airmass at the beginning of the observation was 1.23. A 400\,lines/mm transmission VPH (Volume Phase Holographic) grating is mounted in ACAM and a 1" long slit was used for the experimental design. This configuration yields a resolution R=450 at 5500\,\AA. The seeing during the observation of the target was good, oscillating between 0.6" and 1". The  flux calibration is based on the star SP~0236+052, observed with the same configuration used for the target observation. The wavelength calibration is carried out using the CuNe+CuAr lamps.
The spectrum is also corrected for telluric absorption using a template derived from the standard star spectrum, which was obtained at airmass 1.088. 
The spectrum is corrected for interstellar reddening using $A_V = 0.073$.

\section{Results}
\label{sec:results}

The analysis of the optical spectra from the three gamma-ray sources presented in this work yields to the determination of their redshifts. The uncertainties of the measured redshifts are estimated by performing Monte Carlo simulations of the observed spectrum with the uncertainty estimated from the RMS of the continuum. The spectral features of the simulated spectra are then fitted again with a Gaussian function and the error is obtained from the resulting distribution of centroids.
All of them are inferred from the detection of emission and/or absorption spectral features. While emission lines are found in the spectra from \s\ and \txs, the spectrum from \rs\ is dominated by stellar absorption features. It is consistent with the fact that \rs\ is classified as an extreme gamma-ray blazar, whose synchrotron emission peak is shifted to higher energies w.r.t. other types of blazars, and therefore unveiling the emission from the host galaxy (typically characterized by absorption features). The detailed results on each target are discussed in the following subsections.

\subsection{\s}
\label{sec:results_s}

\begin{figure*}
\includegraphics[scale=0.42]{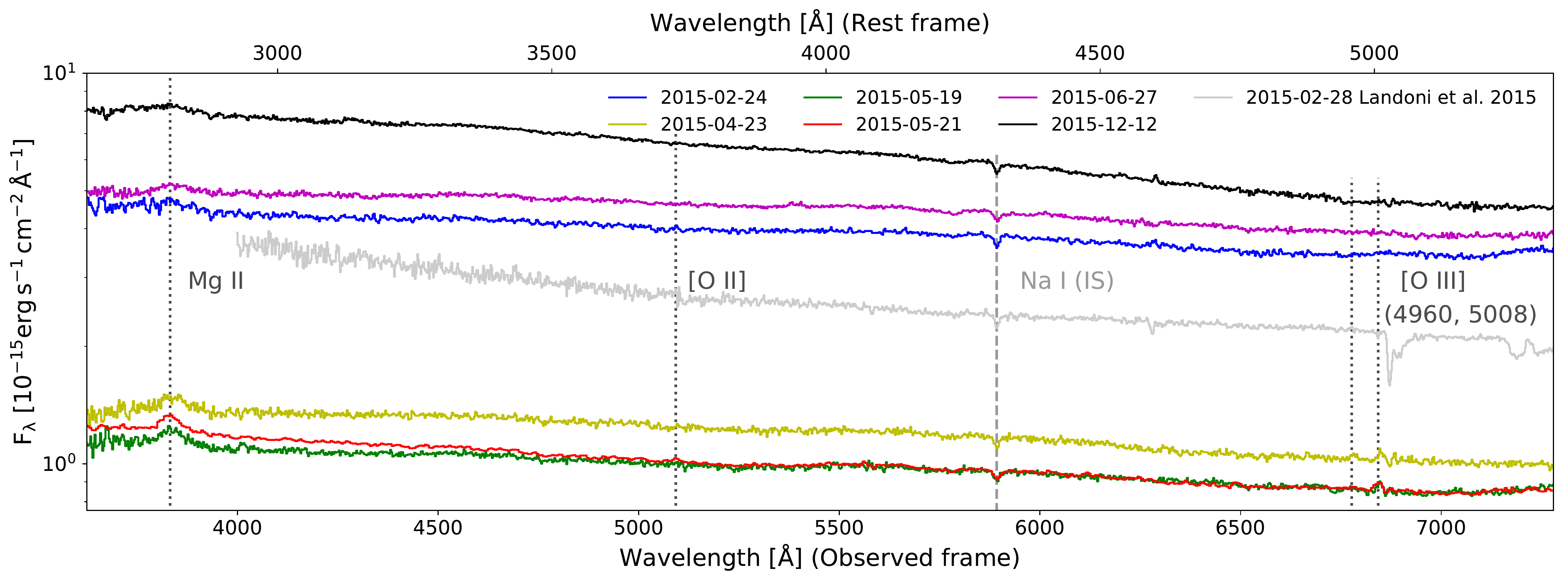}
\includegraphics[scale=0.42]{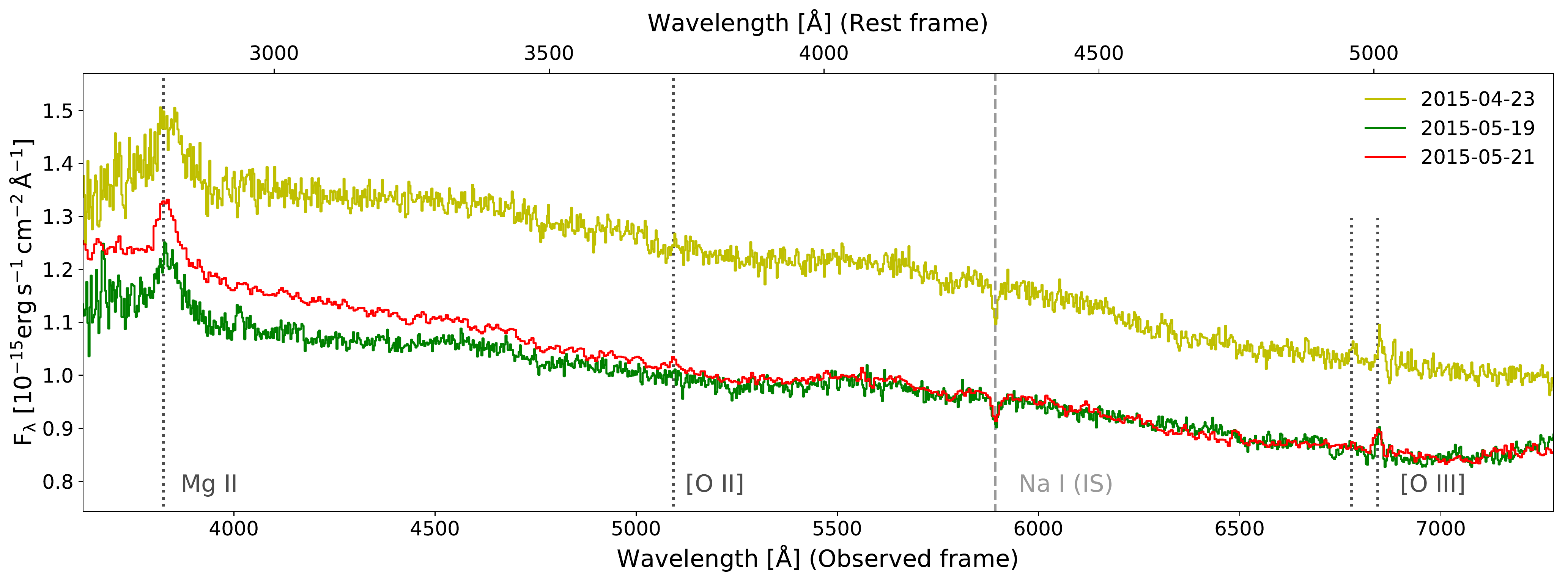}
\caption{\s\ optical spectra and the zoom of the spectra taken during the low state of the source where the spectral features can be identified. The spectrum from \protect\cite{2015AJ....150..181L} is also shown in grey for comparison purposes. All spectra, including the archival one, are flux calibrated using photometric observations (to account for slit losses) and corrected for interstellar reddening. We apply a slit loss correction to the archival GTC spectrum from \protect\cite{2015AJ....150..181L} scaling the flux by the photometric measurement R=15.5 reported in their publication. The spectrum is also corrected for reddening.}
\label{fig:s4}
\end{figure*}

\begin{figure*}
\includegraphics[scale=0.8]{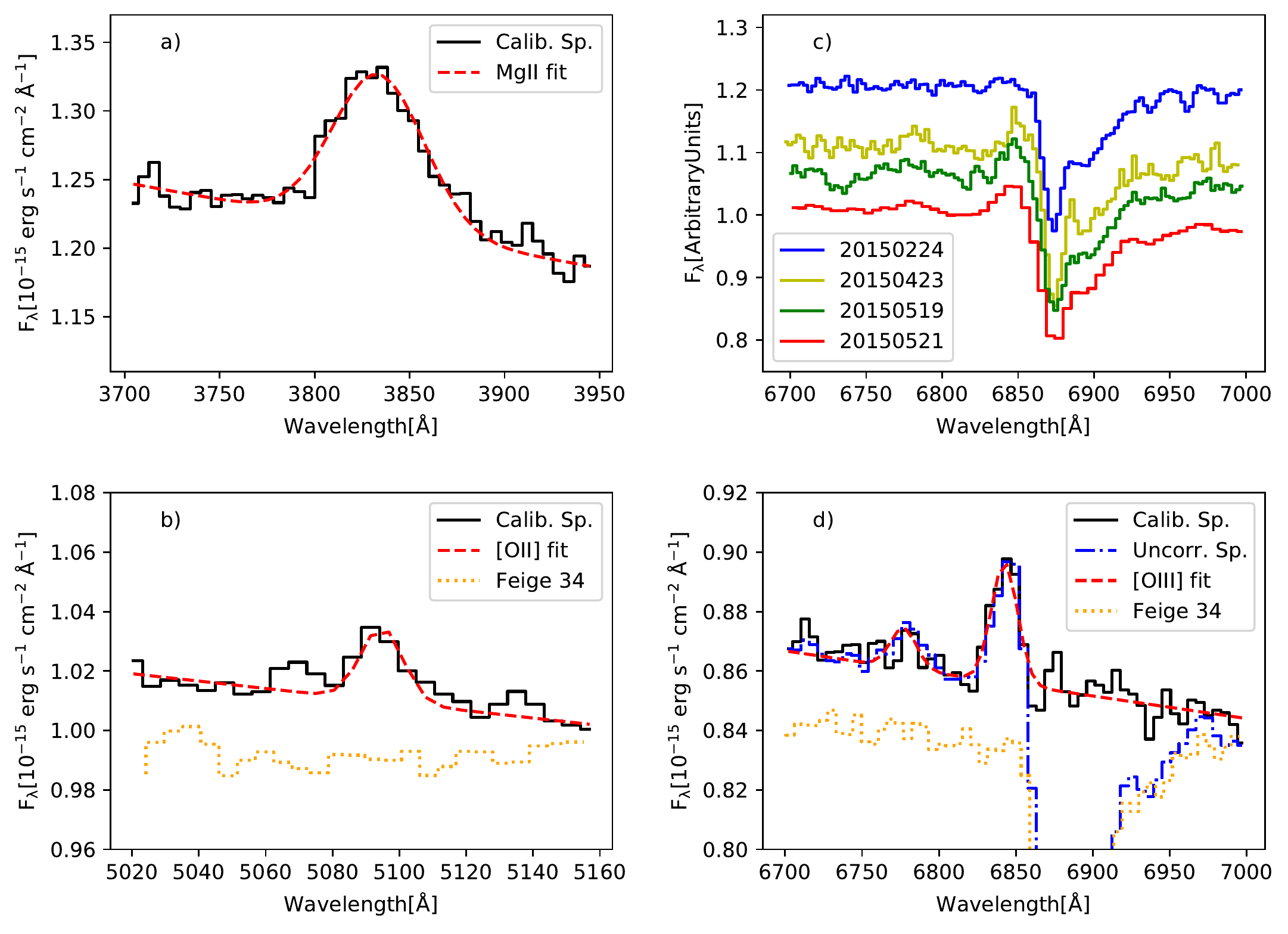}
\caption{Spectral features from \s\ as observed on 2015-05-21, represented in the observed frame. (a) and (b) show the Mg~II and [O~II] lines, respectively. The black solid line represents the flux calibrated spectrum, while the red dashed line denotes the fit. (c) Zoom into the [O~III] region before the telluric correction. The three spectra observed during low state together with one of the high state are shown for comparison purposes. During the low state an excess is visible at the position of the [O~III] 5007\,\AA\ even before the telluric correction is applied, while clearly such excess is not found during the high state. (d) [O~III] region as observed on 2015-05-21. The flux calibrated spectrum before telluric correction is represented by the blue dashed-dotted line, the spectrum after telluric correction is shown by the solid black line and the fit corresponds to the red dashed line. Note that only [O~III] 5007 can be significantly measured as reported in Table~\ref{tab:emission_lines_s4}. The spectrum from the calibration star Feige 34, as observed on 2015-05-21, normalized by the continuum and re-scaled to arbitrary units is shown in panels (b) and (d) for comparison purposes. }
\label{fig:s4_lines}
\end{figure*}

\begin{table*}{Emission Lines Measurements \s\\}
    \centering
    \begin{tabular}{lrrrcc}
       \hline
       \hline
       Line ID  & Center & FWHM & EW$_{\rm{obs}}$ & Line Flux & Line Lum.\\ 
        &  [\AA]  & [$\mathrm{km}\,\mathrm{s}^{-1}$] & [\AA] & [*]
        & [*] 
        \\ 
        \hline
 Mg~II    & 3833.6$\pm$2.7 & 4430$\pm$700 & -4.9$\pm$0.2 & 68$\pm$7 & 31.90$\pm$0.07\\ 
 $\rm{[O~II]3726}$  & 5095.0$\pm$1.3 & 810$\pm$170  & -0.39$\pm$0.07 & 3.5$\pm$0.7 & 1.6$\pm$0.3 \\ 
 $\rm{[O~III]5007}$  & 6842.6$\pm$1.7 & 810$\pm$170  & -0.86$\pm$0.09 & 8.8$\pm$1.2 & 4.1$\pm$0.6\\ 
        \hline 
        \hline
    \end{tabular}
    \caption{Measurement of the optical emission lines from \s. Only the observations performed on 2015-05-21 are considered as it corresponds to the higher S/N spectrum, being the emission lines better characterized. [O~II] line width is fixed to the same value as [O~III] for the line fit. [*] The flux is given in $10^{-16}\,\mathrm{erg}\,\mathrm{cm}^{-2}\,\mathrm{s}^{-1}$ units, while the luminosity is given in $10^{41}\,\mathrm{erg}\,\mathrm{s}^{-1}$.}
    \label{tab:emission_lines_s4}
\end{table*}

The flux calibrated spectra from \s\ taken during different states of the source are shown in Fig.~\ref{fig:s4}. The spectra are highly dominated by the continuum emission from the jet, as typically occurs in blazars. It is specially the case during the high state of the source, where it is difficult to identify any spectral feature except for the Na~I absorption line from the interstellar medium (IS). However, during the low state of the source when the synchrotron radiation from the jet is lower (see zoom plot from Fig.~\ref{fig:s4}), an emission line is clearly visible at the blue end of the spectrum, identified as Mg~II 2800\,\AA. In addition, the [O~III]\ 5007\,\AA\ and a faint [O~II]\ 3726\,\AA\ emission features are present in the spectra at positions consistent with the Mg~II line and previous observations reported in the literature. These features are present only in the three spectra taken during low flux state of the source, namely during April and May 2015. We perform the calculations below using only the best quality spectrum, taken on 2015-05-21 (see S/N ratios in Table \ref{tab:obs_0954}). The emission line flux measurements are obtained by Gaussian fitting and are reported in Table~\ref{tab:emission_lines_s4} for that night. The flux of Mg~II during other nights is consistent with the reported value, while the other lines show low S/N ratios and are not measured. A zoom of the lines and their fits can be found in Fig.~\ref{fig:s4_lines}, where an excess consistent with [O~III] 5007\,\AA\ is clearly visible during low state even before applying the telluric correction. In order to assess the reliability of the feature detections we compute the minimum EW that can be achieved at 3$\sigma$ confidence level (CL), following the procedure described in Appendix~\ref{appendix}.
According to the EW and its uncertainties reported in Table~\ref{tab:emission_lines_s4}, the three emission lines are detected above 5$\sigma$ CL.

There is no sign of stellar absorption features in any of the spectra of \s, which indicates that the host galaxy light is completely outshone by the AGN contribution.
At the measured redshift, the expected host contribution is $\mathrm{F_{\lambda}[6000]} \simeq 3.7 \times 10^{-17} \mathrm{erg\,s^{-1}\,cm^{-2}\,\AA^{-1}}$, which is a factor $\sim 25$ below the lowest flux observed in 2015 May. We assumed $M^{host}_R = -22.5$ \citep{Urry00} and the elliptical template E13 from \cite{2007ApJ...663...81P} to derive such estimation.

We obtain consistent redshift estimates based on the strongest emission lines detected on the spectra taken on 2015-05-21. Thus, the estimated redshift of \s\ is z=$0.3694\pm0.0011$ as derived from the Mg\ II line, and the values z=$0.3667\pm0.0003$ and z=$0.3671\pm0.0003$ from the lines [O~III] and [O~II], respectively. Therefore, our results confirm the previous redshift estimates from \cite{1986AJ.....91..494L, 1996ApJS..107..541L,1993A&AS...98..393S}. Hence, the redshift lower limit of $z\geq0.45$ derived more recently by \cite{2015AJ....150..181L} is ruled out based on the results on this work. As shown in Fig.~\ref{fig:s4}, the redshift lower limit was estimated based on a featureless spectrum while the source was still in flaring state and the  wavelength coverage does not include the prominent Mg~II line. As estimated in Appendix~\ref{appendix}, the minimum EW ($\mathrm{EW}_\mathrm{min}$) at $3\,\sigma$ CL is 0.23\,\AA\ for OSIRIS/GTC spectrum comparable to the 0.26\,\AA\ calculated for the DOLORES/TNG spectrum. However, the expected EW of the [O~III] emission line should be scaled down by a factor $\simeq3$ to account for the flux difference between the two spectra. It results in a value 0.29 which is of the order of the $\mathrm{EW}_\mathrm{min}$ based on the measured value in the 2015 May spectrum (Table~\ref{tab:emission_lines_s4}). This would explain the vanishing of the emission line features observed in the OSIRIS/GTC spectrum (see Appendix~\ref{appendix_landoni}).

We have estimate the black-hole mass using a scaling relation valid for the Mg~II line. We used the expression provided by \citet{VestergaardOsmer09} which relates the line width and the luminosity of the ionizing continuum, identified with the big blue bump. In order to estimate this contribution and distinguish it from the jet emission, we use the relationship between the predicted luminosity at 3000 \AA\ ($\mathrm{L}_{\mathit{pred}}$) and the Mg~II line luminosity, as given by \citet{Shen11}. From this value we can also estimate the parameter NTD \citep[Non-thermal Dominance, which is the ratio between observed and predicted luminosities,][]{2012ApJ...748...49S}, resulting in a value NTD=6.4, or equivalently a value of $\mathrm{L}_{\mathit{pred}}=1.8 \times 10^{-16} \mathrm{erg}\,\mathrm{cm}^{-2}\,\mathrm{s}^{-1}\,\mathrm{\AA}^{-1}$. Using this luminosity and the FWHM of Mg~II line we obtain a $M_{BH} \simeq 2.3 \times 10^8 M_\odot$, with an uncertainty of a factor 3.5 as given by the scaling relationship.

The luminosity of the line [O~III] in \s\ is among the less luminous QSO contained in the SDSS \citep{Shen11}. Furthermore, the line ratio $\mathrm{[O~III]/[O~II]}=2.3\pm0.5$ indicates high ionization gas, whose excitation is likely due to photoionization by the nuclear emission \citep{MoyRoccaVolmerange02}. 

Regarding the classification of the source, according to the EW at rest frame measure for the Mg~II line, it is $< 5$\,\AA\ (see Table~\ref{tab:emission_lines_s4}) and therefore, the source can be classified as a BL~Lac object. However, it is close to the limit of 5 \AA, possibly being consistent with the existence of a weak BLR. From the MWL SEDs of \s\ reported in \cite{2016PASJ...68...51T,2018A&A...617A..30M}, the high CD reported in both works, clearly demonstrates that standard one-zone SSC models fail to explain the MWL emission from the source. Indeed, in order to successfully explain the MWL emission from \s\ external photon fields are invoked in addition to the SSC emission. In particular, to avoid absorption of gamma rays (specially in the VHE band) within a possible BLR, the authors assume a location of the emitting region closer to a possible IR torus and use such external photon field as the possible source of radiation external to the jet. Therefore, this source could be classified as a transitional blazar: on the one hand it displays relatively weak emission lines and on the other hand, the CD is similar to the one usually present in FSRQs (requiring the same type of theoretical modeling including external photon fields).

From the optical spectra presented in this work we can derive the expected luminosity of the BLR and the torus in order to compare it with the photon field needed to explain its MWL SED. On the first place, the luminosity of the BLR can be estimated as proposed by \citet{1997MNRAS.286..415C}, $L_{\rm BLR} \simeq 5.6 L_{Ly\alpha}$. Since the only broad line detected in our spectra is Mg~II, we use the average relative emission line ratio between Mg~II and Ly$\alpha$ as 34\% as reported by \citet{1991ApJ...373..465F}. Therefore, the relation between the luminosity of the BLR and the luminosity of the Mg ~I emission line should be $L_{\rm BLR} \simeq 16.5\,L_{\rm Mg II} \simeq 5.3 \cdot 10^{43}\,\mathrm{erg}\,\mathrm{s}^{-1}$. This result can be cross checked using an updated line ratios estimation from the SDSS reported by \cite{2001AJ....122..549V}. Following the same strategy as in \cite{1997MNRAS.286..415C}, the sum of all the emission line ratios (w.r.t. $Ly\alpha=100$) results in $L_{\rm BLR} \simeq 263.87 \simeq 2.64 L_{Ly\alpha}$. According to this work, the Mg~II line is 14.725\% of $Ly\alpha$, yielding $L_{\rm BLR} \simeq 5.7 \cdot 10^{43}\,\mathrm{erg}\,\mathrm{s}^{-1}$. Both results are consistent within an 8\% uncertainty. As it is the standard recipe generally used, the rest of the calculations are based on the result from \citet{1997MNRAS.286..415C} and \citet{1991ApJ...373..465F}.

From the estimation of the $L_{\rm BLR}$ and assuming a standard covering factor of the BLR of $\rm{f}_{\rm cov_{BLR}}=0.1$ \citep[maximum value found is 0.15 by][]{1981MNRAS.195..437S}, an estimation of the disk luminosity can be derived as $L_{\rm disk} \approx 1/f_{\rm cov_{BLR}}\cdot L_{\rm BLR}\approx 5.3 \cdot 10^{44}\,\mathrm{erg}\,\mathrm{s}^{-1}$. The estimated $L_{\rm disk}$ is consistent with the distribution found by \cite{2014Natur.515..376G}. The IR torus luminosity can also be derived from the disk luminosity, assuming a certain covering factor. The standard values for the covering factor for the torus range from 0.5 to 0.7 according to \cite{2012MNRAS.425L..41C}. In this work a mean value of 0.6 will be used. Therefore, the torus luminosity for \s\ is estimated as $L_{\rm torus}=f_{\rm cov_{torus}}\cdot L_{\rm disk} \simeq 3.2 \cdot 10^{44}\,\mathrm{erg}\,\mathrm{s}^{-1}$. The radius of the IR torus can be inferred as $r_{\rm torus}\approx 2.5\cdot 10^{18} \,[L_{\rm disk}/(10^{45} \, {\rm erg} \, {\rm s}^{-1})]^{1/2} \,{\rm cm} \simeq 1.8 \cdot 10^{18} \,\rm{cm} \simeq 0.6  \,\rm{pc}$ as prescribed by \cite{2009MNRAS.397..985G}.

From the MWL SED modeling of \s\ in \cite{2018A&A...617A..30M}, the physical environment responsible for the external IR photon field is assumed as $L_{\rm torus}=1.5 \cdot 10^{42} \mathrm{erg}\,\mathrm{s}^{-1}$ and $r_{\rm torus}=6.1 \cdot 10^{17} \rm{cm}=0.2\,\rm{pc}$. Therefore the assumed IR torus luminosity is more than two orders of magnitude below the luminosity compatible with the observed Mg~II emission line in its optical spectrum, while the assumed radius of the torus is around 3 times smaller than the dimension derived in this work. From these results we can conclude that even if the source is classified as a BL~Lac object, its optical spectrum is compatible with the existence of an IR torus powerful enough to provide the external photon field required for the production of gamma rays needed to successfully explain its MWL SED.

\subsection{\txs}

\begin{figure*}
\includegraphics[scale=0.42]{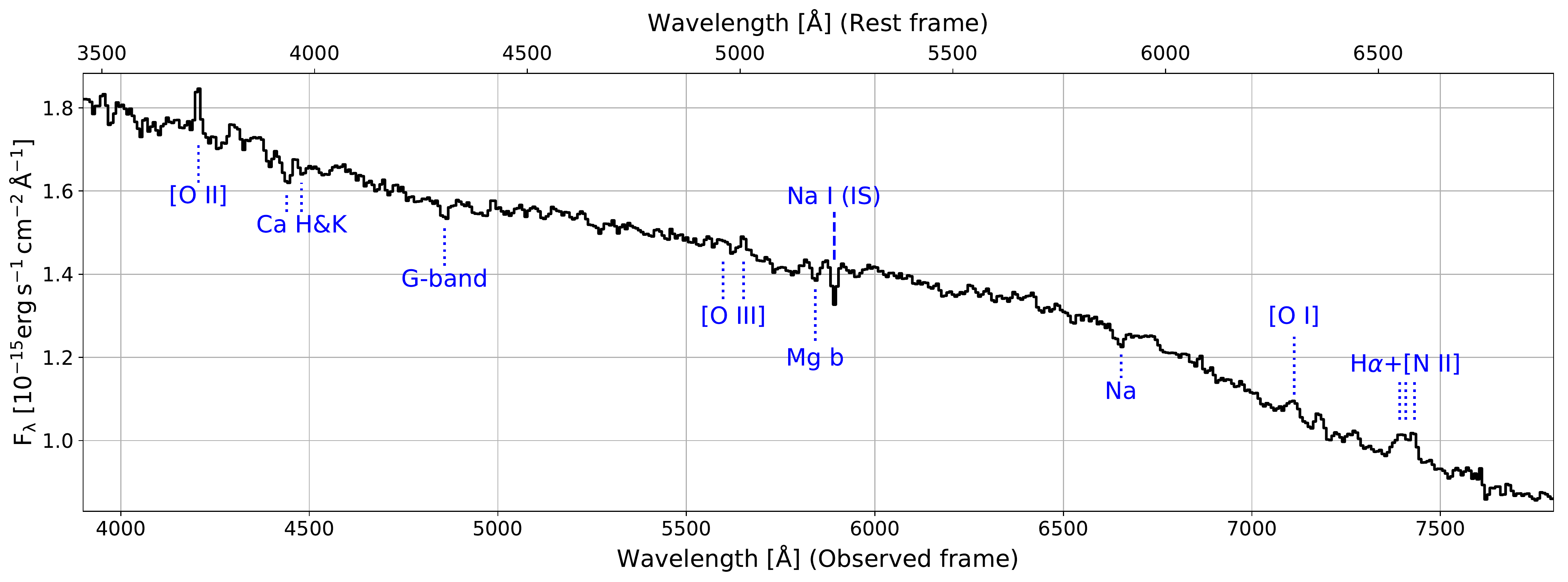}
\caption{\txs\ optical spectrum obtained at NOT on 2019 May 7. The detected spectral features are consistent with a redshift value of $z=~0.1281~\pm~0.0004$. See text for details.}
\label{fig:txs}
\end{figure*}

\begin{figure*}
\includegraphics[scale=0.8]{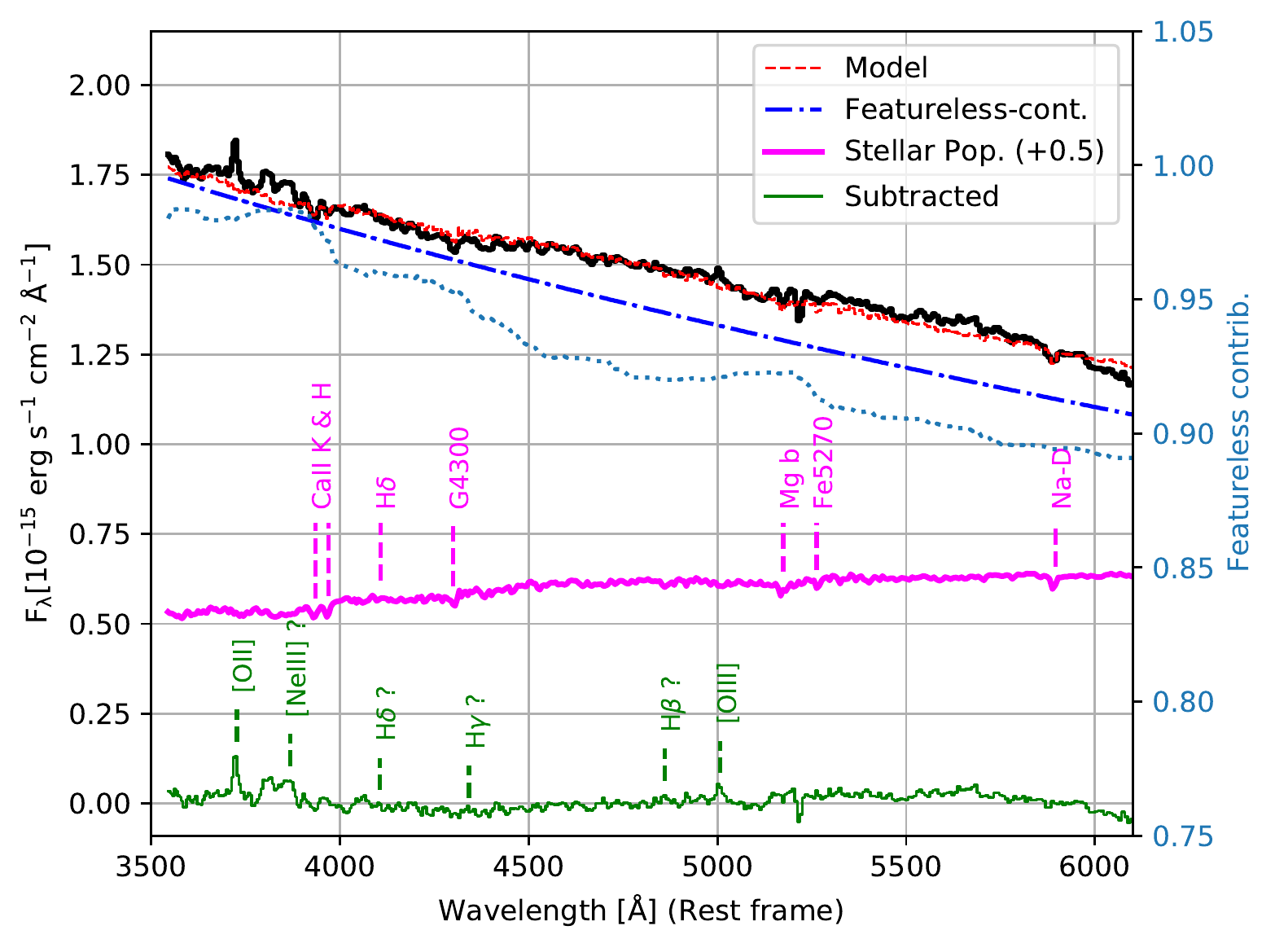}
\caption{Optical spectra of \txs\ (black) and best fit model (red) obtained using pPXF, which includes a featureless continuum (blue) and the stellar population contribution (magenta), adding an offset of $0.5$ flux units for representation purpose. 
The subtraction of the model to the observed spectra is shown at residual level (green line). It can be noticed that the stellar absorption lines are relatively well reproduced and the emission lines emerges. Absorption and emission spectral features are marked in the figure. The right y-axis refers to the contribution of the featureless continuum to the observed spectrum (dotted blue line). 
}
\label{fig:txs_ppxf}
\end{figure*}

The flux calibrated spectrum from \txs\ is measured with a S/N of 227 in the range 4500-4800\,\AA\ (see Fig.~\ref{fig:txs}). Both emission and absorption spectral features can be identified. The list of the detected spectral features is reported in Table~\ref{tab:spectral_lines_txt}. The derived redshift from such spectral features is $z~=~0.1281\pm0.0004$, whose uncertainty is estimated from the [O~II] emission line. The minimum EW at $3\,\sigma$ CL is derived for a couple of spectral regions, in the blue for the G-band resulting in $\rm{EW}_{min}$=0.31\,\AA\, and in the red part around the Na line yielding $\rm{EW}_{min}$=0.28\,\AA (see Fig \ref{fig:EWs_S4_TXS}). The redshift determination is secured thanks to the detection of several spectral lines with a high CL, only few features reported in Fig.~\ref{fig:txs} and Table~\ref{tab:spectral_lines_txt} are below the $3\,\sigma$ CL as Ca\,K\&H for instance.

Due to the fact that only weak emission lines are detected from \txs\ with small equivalent widths ($\rm{EW}_{\rm{rest}}<5$\AA), the target can be classified as a BL~Lac object. The emission line ratios [O~III]/[O~II]$\simeq 0.5 \pm 0.17$ and [NII]/$\rm{H}{\alpha} \simeq 1.7 \pm 0.9$ indicate a very low ionization level of the line--emitting gas, as observed in LINERS or Low-Luminosity AGNs (LLAGNs). The most plausible excitation mechanism in this case is a combination of photoionization and shocks \citep{MoyRoccaVolmerange02}. The luminosity of the line [O~III] is relatively low ($log(\rm{Lum([O~III])=}40.7$) which is compatible with the nucleus being a LLAGNs. There are predictions that at low nuclear luminosities the BLR should disappear \citep{ElitzurHo09},  which is consistent with the non-detection of broad lines in the spectra of \txs. 

The presence of weak stellar absorption features in the spectrum of \txs\ suggests that the stellar population contribution is significantly diluted by the nuclear non-stellar continuum.
In order to uncover the underlying stellar population we apply the penalized pixel fitting technique (pPXF) \citep{Cappellari17}. We use the same set of synthesis models as in \citet{2020MNRAS.494.6036B}, where a detailed description can be found.  
The regions showing emission lines and the Na\,I interstellar absorption feature are masked prior to the modelling. Given the dominance of the non-stellar contribution we model the observed optical spectrum by adding to the stellar templates a second order Legendre polynomial (in a log($\lambda$) scale), which permits to improve the fit without varying the template continuum shape. The addition of this polynomial component mimics the commonly observed power law component and accounts for the likely featureless contribution from the active nucleus. The mass-weighted age and metallicity of the best fitting models indicates a very old ($<\log(\rm{Age})>= 1.112 \pm 0.002$\,Ga) and metallic ($<[M/H]>=0.25\pm0.04$) stellar population.  

\citet{Urry00} claimed that BL~Lac objects are hosted by giant elliptical galaxies showing a narrow distribution of $M^{\mathit{host}}_R$, independently of the nuclear luminosity. 
The contribution of the stellar population from the host galaxy can be obtained from the pPXF modelling. We estimate the host contribution to the flux within the slit in the $R$-band around 10\,\% (see Fig. \ref{fig:txs_ppxf}).
We estimate around 1.7 mags slit losses due to the extension of the host galaxy assuming a de Vaucoleurs profile (Sersic n=4) for $r_{\mathit{eff}}=7\,\mathrm{kpc}$ \cite[compatible with][in their Fig. 3 at the estimated redshift]{Urry00}, corresponding to $r_{\mathit{eff}} = 2.97\arcsec$.
Thus the apparent magnitude in R of the stellar population is $m^{host}_R \sim 16.3$, including interstellar reddening correction.

 From these values and the derived redshift we can estimate the absolute magnitude of the host galaxy, which results in $M^{\mathit{host}}_R \rm -22.6$, which is just above the peak of the brightness distribution of BL~Lac host absolute magnitudes according to \citet{Shaw13}, but compatible within the uncertainties given in that work ($M^{host}_R= -22.5 \pm 0.5$). The derived values indicate that the host galaxy should be a luminous bright  elliptical galaxy. A more accurate determination of the host brightness should be carried out after a morphological study using deep high-resolution imaging.

\begin{table}[Spectral Features \txs]
    \centering
    \begin{tabular}{lrrrc}
       \hline
       \hline
       Line ID  & Center & EW$_{\rm{obs}}$ & Flux & Lum. \\ 
     &   [\AA]  & [\AA] & [*] & [*] 
     \\
        \hline
     $\rm{[O~II] 3727}$ & 4204.3$\pm$1.5 & -1.42$\pm$0.26 & 2.12$\pm$0.36 & 9.25$\pm$1.58\\ 
     Ca K & 4440$\pm$4 &  1.3$\pm$0.5 & ... & ... \\ 
     Ca H & 4484$\pm$7 &  0.7$\pm$0.5 & ... & ... \\ 
     G-band    & 4859$\pm$1 &  0.47$\pm$0.10 & ... & ... \\
     $\rm{[O~III] 5007}$    & 5651$\pm$4 &  -1.03$\pm$0.29 & 1.16$\pm$0.31 & 5.08$\pm$1.34\\ 
     Mg b    & 5842$\pm$5 &  0.70$\pm$0.15 & ... & ... \\ 
     Na    & 6651$\pm$4 &  1.04$\pm$0.10 & ... & ... \\ 
     $\rm{[O~I] 6300}$    & 7110$\pm$5 &  -1.26$\pm$0.35 & 1.17$\pm$0.31 & 5.13$\pm$1.36\\ 
     H${\alpha}^\dagger$   & 7401$\pm$4 & -3.8$\pm$0.6 & 1.06$\pm$0.50  & 4.65$\pm$2.18 \\
     $\rm{[N~II] 6584}^\dagger$   & 7424$\pm$4 & -3.8$\pm$0.6 & 1.81$\pm$0.41  & 7.94$\pm$1.78\\     
        \hline 
        \hline
    \end{tabular}
    \caption{Spectral features detected in the optical spectrum from \txs. [*] The flux is given in $10^{-15}\,\mathrm{erg}\,\mathrm{cm}^{-2}\,\mathrm{s}^{-1}$ units, while the luminosity is given in $10^{40}\,\mathrm{erg}\,\mathrm{s}^{-1}$. Notes: $^\dagger$The center and width of the lines [N~II]~6548,6584 are linked to the values corresponding to H$\alpha$. The line flux ratio [N~II]~6584/6548 is kept fixed to the theoretical value 3. The EW includes the lines H$\alpha$ and [NII]~6548,6584.}
    \label{tab:spectral_lines_txt}
\end{table}

\subsection{\rs}

\begin{figure*}
\includegraphics[scale=0.42]{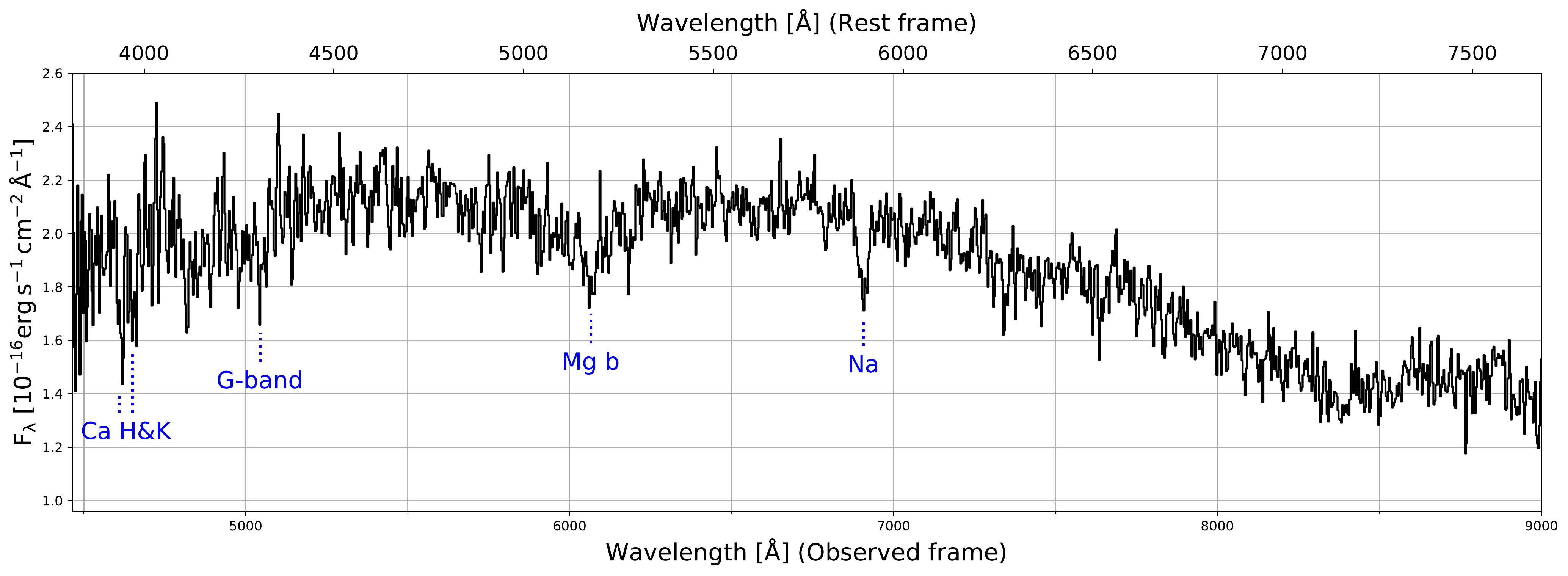}
\caption{\rs\ optical spectrum observed at WHT on 2017 October 20. Based on the detection of its absorption spectral features, a redshift of $z=0.1721\pm 0.0002$ is derived. See text for details.}
\label{fig:rs}
\end{figure*}

\begin{table}[Absorption Features \rs]
    \centering
    \begin{tabular}{lrr}
       \hline
       \hline
       Line ID  & Center &  EW$_{\rm{obs}}$ \\ 
     &   [\AA]  & [\AA]  
     \\
        \hline
     Ca K    & 4615$^{*}$ &  4.0$^{*}$  \\ 
     Ca H    & 4652$^{*}$ &  4.8$^{*}$  \\ 
     Mg b    & 6064.9$^{*}$ & 5.5$^{*}$ \\ 
     Na    & 6906.1$\pm$1.4 & 5.4$\pm$0.5 \\ 
        \hline 
        \hline
    \end{tabular}
    \caption{Absorption lines detected in the optical spectrum from \rs. $^*$ These features have been measured in the smoothed spectra, hence their uncertainties  cannot be estimated using our method based on Monte Carlo simulations.}
    \label{tab:spectral_lines_rs}
\end{table}

The flux calibrated optical spectrum from \rs\ is shown in Fig.~\ref{fig:rs}. Although the S/N is only 17 (calculated in the range 5000-5500\,\AA\,) it can be clearly appreciated that the spectrum is dominated by the emission from the host galaxy, and therefore, the absorption features dominate the spectrum. The different absorption lines identified are listed in Table~\ref{tab:spectral_lines_rs}. The corresponding estimated redshift is $z=0.1721\pm 0.0002$, based on the Na absorption line. 

The observed spectrum from \rs\ is compatible with the classification as extreme blazar from the 2WHSP catalog \citep{2017A&A...598A..17C}, as its optical emission is dominated by the host galaxy and not by the emission from the relativistic jet. This fact indicates that the emission peaks from the MWL SED are located at high energies, unveiling the emission from the host galaxy instead of outshining it as it is commonly observed in less energetic blazars. 

\begin{figure*}
\includegraphics[scale=0.8]{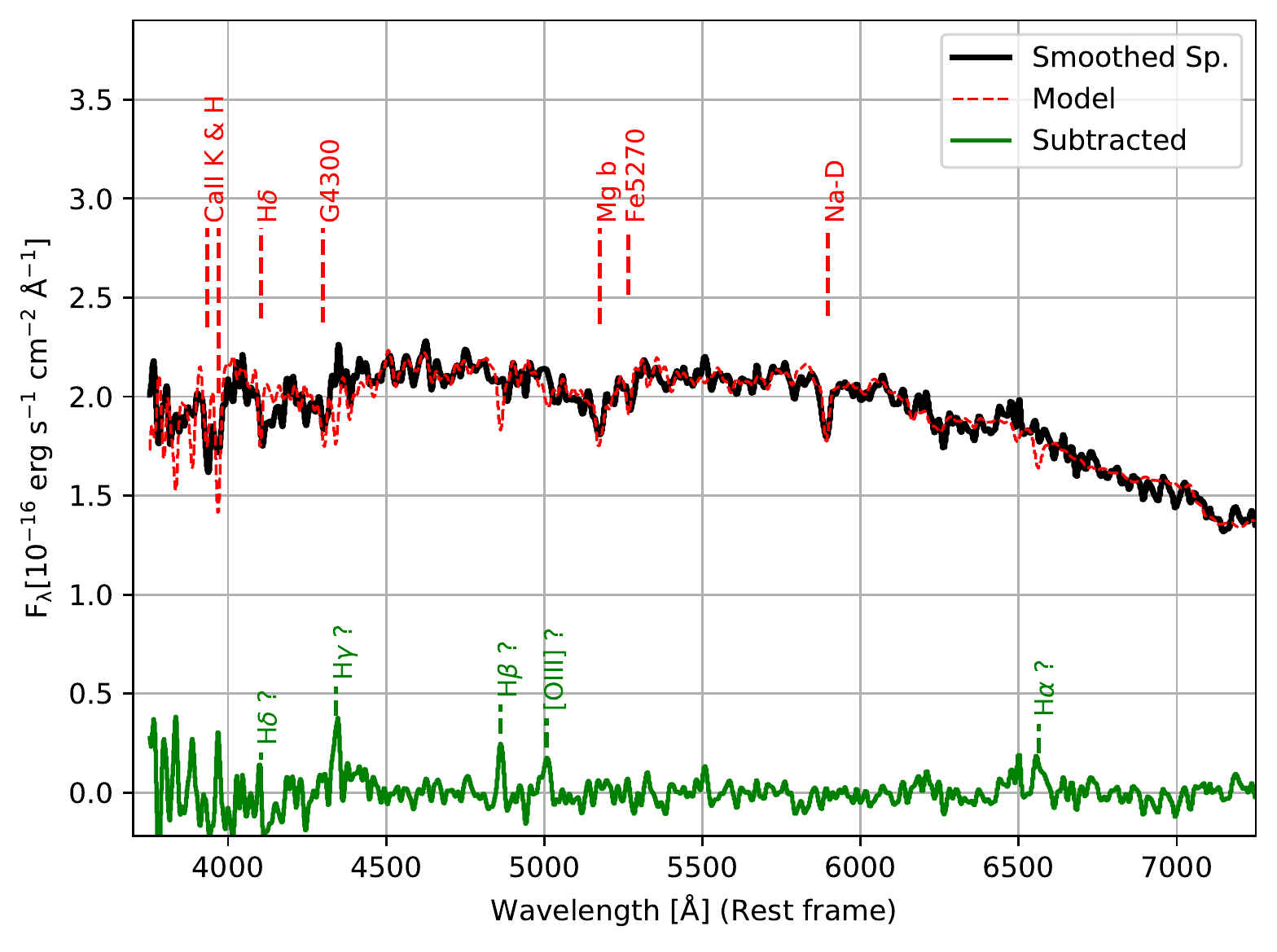}
\caption{Smoothed optical spectra of \rs\ (black) and best fit model (red) obtained using pPXF. 
The subtraction of the model to the observed spectra is shown at residual level (green line). It can be noticed that the stellar absorption lines are relatively well reproduced and possible emission lines emerge. Absorption and emission spectral features are marked in the figure.}
\label{fig:rs_ppxf}
\end{figure*}

Given the high noise level observed in the spectrum we apply a Gaussian filter (FWHM=4 pixels, as corresponding to the slit width). 
We try to model the stellar population using pPXF to reproduce the smoothed optical spectrum of \rs.  We cannot firmly rely on the resulting spectral shape given the limited S/N ratio of our spectrum. In order to improve the modelling we use a multiplicative polynomial function to adjust the continuum shape of the template to the measured spectrum. The best results are obtained with an order 3 polynomial which accounts for deviations of the spectral shape of $\pm 10$\% with respect to the synthesis templates. 
The best fit is obtained for a model with weighted-mass age and metallicity values of  $<\log(\rm{Age})>= 1.059 \pm 0.003$\,Ga and $<[M/H]>=0.183\pm0.011$, respectively. 


Our model fit confirms that the optical emission from \rs\ is completely dominated by the host stellar population. 
Hence, the host galaxy luminosity at the $R$-band can be estimated from the photometry reported in the SDSS. 
We have used the value $r=16.85$ obtained from the cmodel\footnote{\url{https://www.sdss.org/dr12/algorithms/magnitudes/}} and assumed it as $R$-band. Note that the magnitude reported from the PSF modelling ($r=17.629$) is in better agreement with the flux level of the observed spectrum. These results can be explained if the host galaxy is resolved for seeing limited observations, which is compatible with the difference between cmodel and PSF values reported in the SDSS archive. The luminosity distance is computed using the cosmological calculator by \citet{2006PASP..118.1711W}. Thus the host absolute magnitude in the $R$-band is $\sim -22.8$, which is close to the maximum of the BL~Lacs host brightness distribution as found by \citet{Shaw13}. The derived value is consistent with the host galaxy of \rs\ being a large elliptical galaxy, as expected for a bright high-energy BL~Lac.

\section{Conclusions}
\label{sec:conclusions}

In this work, we present the redshift estimation of three gamma-ray blazars whose distance was previously either unknown or under debate. Moreover, the optical spectroscopic information from those sources are also used to classify them. The individual conclusions from the study of each target are:
\begin{itemize}
    \item{\s: a solid redshift determination of z=$~0.3694~\pm~0.0011$ is established for this target, confirming the historical measurements from \cite{1986AJ.....91..494L, 1996ApJS..107..541L,1993A&AS...98..393S} and ruling out the lower limit of $z\geq0.45$ derived by \cite{2015AJ....150..181L} on February 2015. 
    This work shows the importance of the state of the source during the optical spectroscopic observations. When the source is in high state the continuum emission from relativistic jets can easily outshine the spectral features. During the observations performed by \cite{2015AJ....150..181L} the source displayed a higher flux (around 3 times) than the spectra in low state presented in this work and the historical observations. This makes more difficult to detect the [O~III] line very close to a telluric absorption feature. In addition, the strong emission line Mg~II detected in the DOLORES/TNG spectra is located at $\sim$\,3830\,\AA\ (see Fig.~\ref{fig:s4}), a wavelength range which is not covered in the OSIRIS/GTC spectrum.
    With this work, we can also confirm the classification of \s\ as a BL~Lac object, but close to the limit and therefore, it seems to confirm that the target could be a transitional object showing characteristics from both BL~Lacs and FSRQs. 
    The line ratio [O~III]/[O~II] indicates photoionization by a hard continuum such as the expected from the active nucleus.
    From the observed Mg~II emission line the luminosity of the disk, BLR and torus is derived assuming commonly used values for the covering factors.
    The results demonstrate that the observed optical spectrum is compatible with the existence of a torus luminous enough to provide the IR external photon field needed for the production of gamma rays as required by the high CD MWL SED. A black-hole mass of $M_{BH} \simeq 2.3 \times 10^8 M_\odot$ is estimated.}
    
    \item{\txs: the redshift determination of the source as $z=0.1281\pm0.0004$ is presented in this work, being this information crucial for further studies of its gamma-ray emission. The source classification as a BL~Lac object is confirmed based on the weak optical emission lines detected. 
    The line ratios [O~III]/[O~II] and [N~II]/H$\alpha$ are compatible with gas excitation produced by shock heating.
    While its optical spectrum is dominated by the continuum emission from the jet, the weak underlying stellar population is modeled using the pPXF technique. The best fit found yields an old and metallic stellar population with $<\log(\rm{Age})>= 1.112 \pm 0.002$\,Ga and $<[M/H]>=0.25\pm0.04$. The absolute magnitude of the host galaxy is estimated as $M^{\mathit{host}}_R \simeq -22.6$, comparable with the peak of the luminosity distribution of the BL~Lac sample from \cite{Shaw13}.}

    \item{\rs: from the spectral shape and the detection of absorption features we can confirm that the spectrum is dominated by its host galaxy, as typically occurs in extreme blazars. Therefore, the results presented in this work are consistent with the classification of the source as a 2WHSP target. Its redshift can be firmly established for the first time as $z=0.1721\pm 0.0002$, thanks to the identification of a few optical spectral features. The stellar population is modeled with pPXF, yielding $<\log(\rm{Age})>= 1.059 \pm 0.003$\,Ga and $<[M/H]>=0.183\pm0.011$. The analysis results in an absolute magnitude of the host galaxy of $M^{\mathit{host}}_R\sim -22.8$, again close to the average magnitude found by \cite{Shaw13} for a BL~Lac sample.} 
    
\end{itemize}

All this new information is essential for the study of the intrinsic characteristics of the three studied gamma-ray BL~Lac objects, since the redshift is a key information to determine the EBL absorption suffered by the gamma-ray spectrum (specially in the VHE band). Moreover, the classification of the target and the study of the characteristics of a possible BLR is also important information for the modeling of their MWL SEDs. In addition to the use of our results for the study of these sources within the current gamma-ray context, they will be better studied using the capabilities of the next generation of Cherenkov telescopes, the Cherenkov Telescope Array (CTA) currently under construction.


\section{Acknowledgements}
We thank Paolo Goldoni, Roopesh Ojha and Tapio Pursimo for useful discussions. We thank Elina Lindfors and the KVA team for providing information on the optical magnitude of \txs.
JBG acknowledges the support of the Viera y Clavijo program funded by ACIISI and ULL. JBG and JAP acknowledges financial support from the Spanish Ministry of Economy and Competitiveness (MINECO) under the 2015 Severo Ochoa Program MINECO SEV-2015-0548. JOS thanks the support from grant FPI-SO from the Spanish Ministry of Economy and Competitiveness (MINECO) (research project SEV-2015-0548-17-3 and predoctoral contract BES-2017-082171).

Based on observations made with the Italian Telescopio Nazionale Galileo, the Nordic Optical Telescope and the William Herschel Telescope. These telescopes are located in the Spanish Observatorio del Roque de los Muchachos of the Instituto de Astrofisica de Canarias, and are operated by the Fundación Galileo Galilei of the INAF (Istituto Nazionale di Astrofisica), the Nordic Optical Telescope Scientific Association and the Isaac Newton Group, respectively.

\section{Data Availability}
The data reported in this work are available at the archival databases from the different telescopes used: TNG, WHT and NOT.



\bibliographystyle{mnras}
\bibliography{bibliography} 

\appendix

\section{Estimation of the minimum EW in featureless or faint-features spectra}
\label{appendix}

A practical way to evaluate whether a spectral feature is detected on a given spectrum is to compare the measured EW with the minimum measurable value ($\rm{EW}_{min}$). If no feature is present, the measurements of the EW within a wavelength interval will produce a distribution centered around zero, and its width is related to the noise of the spectrum.  The detection of a spectral feature is assumed to be significant when its EW exceeds 3 times the standard deviation of the distribution.    
This approach was previously proposed by \citet{Sbarufatti05,2015AJ....150..181L,2017ApJ...837..144P} as a tool to provide lower limits to the redshift in BL Lac objects showing a featureless spectra.

Here we propose to compute the $\rm{EW}_{min}$ from the noise observed in the continuum adjacent to a given feature position. In practice, the continuum is fitted with a low order polynomial (commonly a straight line or a parabola when curvature is evident within the wavelength interval) and subtracted to the observed spectrum. The continuum noise ($\mathit{RMS}_{c}$) is obtained as the standard deviation from the residuals. Here it is assumed that the fluctuations between adjacent pixels are uncorrelated, which is a good approach for faint source spectra and/or high-to-moderate resolution spectra. Once the noise-per-pixel is known, the minimum EW can be computed by the expression: 
$$\mathrm{EW}_{min} = n_\sigma \, \sigma(\mathrm{EW}) \simeq n_\sigma \, \frac{\sqrt{N_{pix}} \mathit{RMS}_{c}  \delta\lambda}{F_c(\lambda)},$$ 
where $N_{pix}$ is the number of pixels within the wavelength interval containing the spectral feature and $\delta\lambda$ is the spectral dispersion. 
This expression can also be written in terms of the signal-to-noise ratio of the continuum ($\mathrm{SNR}_c$) 
and the wavelength window ($\Delta\lambda$) as: 
$$\mathrm{EW}_{min} \simeq n_\sigma \, \frac{\sqrt{\Delta\lambda \, \delta\lambda}}{\mathrm{SNR}_c},$$ 

\begin{figure*}
\includegraphics[scale=0.5]{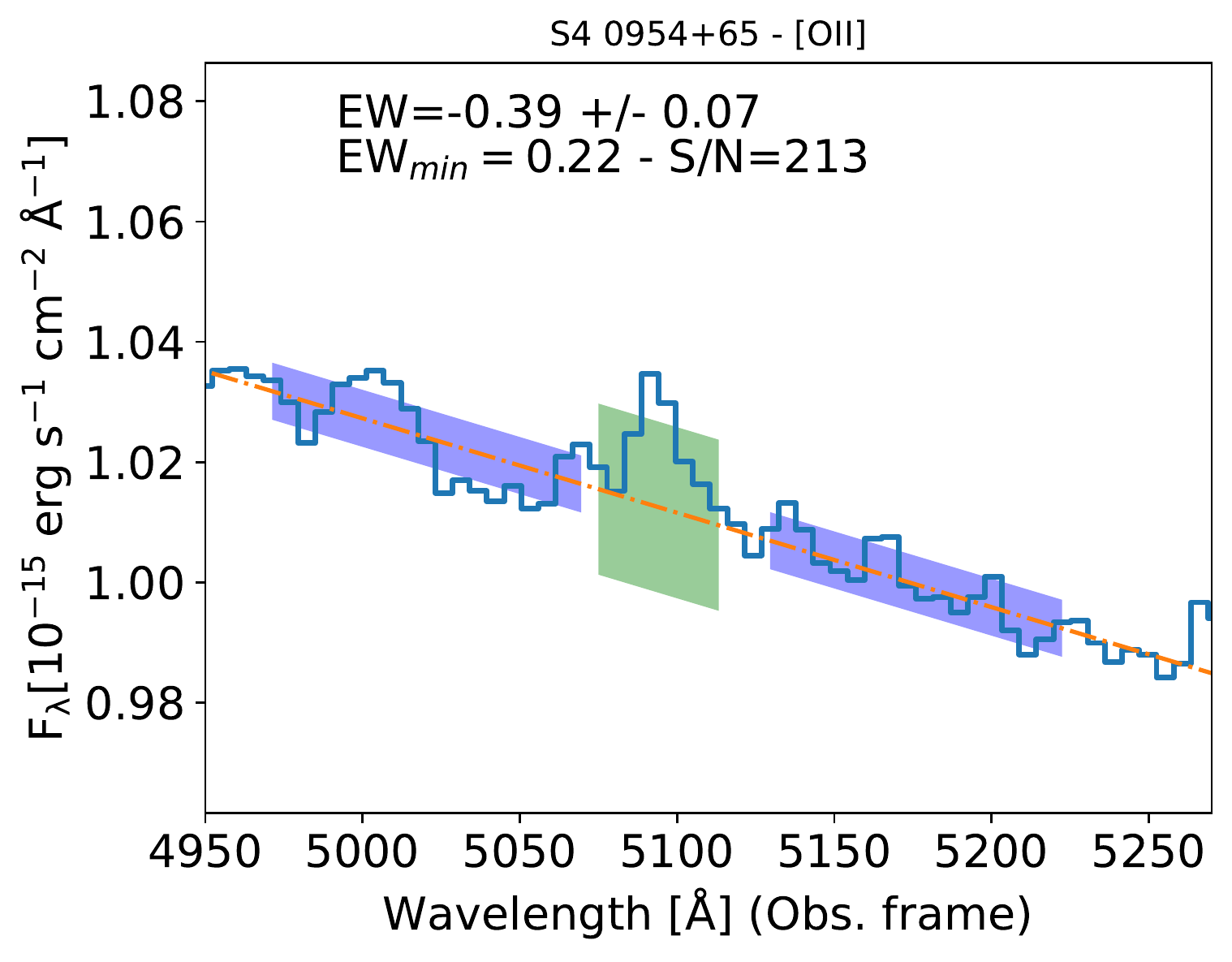}
\includegraphics[scale=0.5]{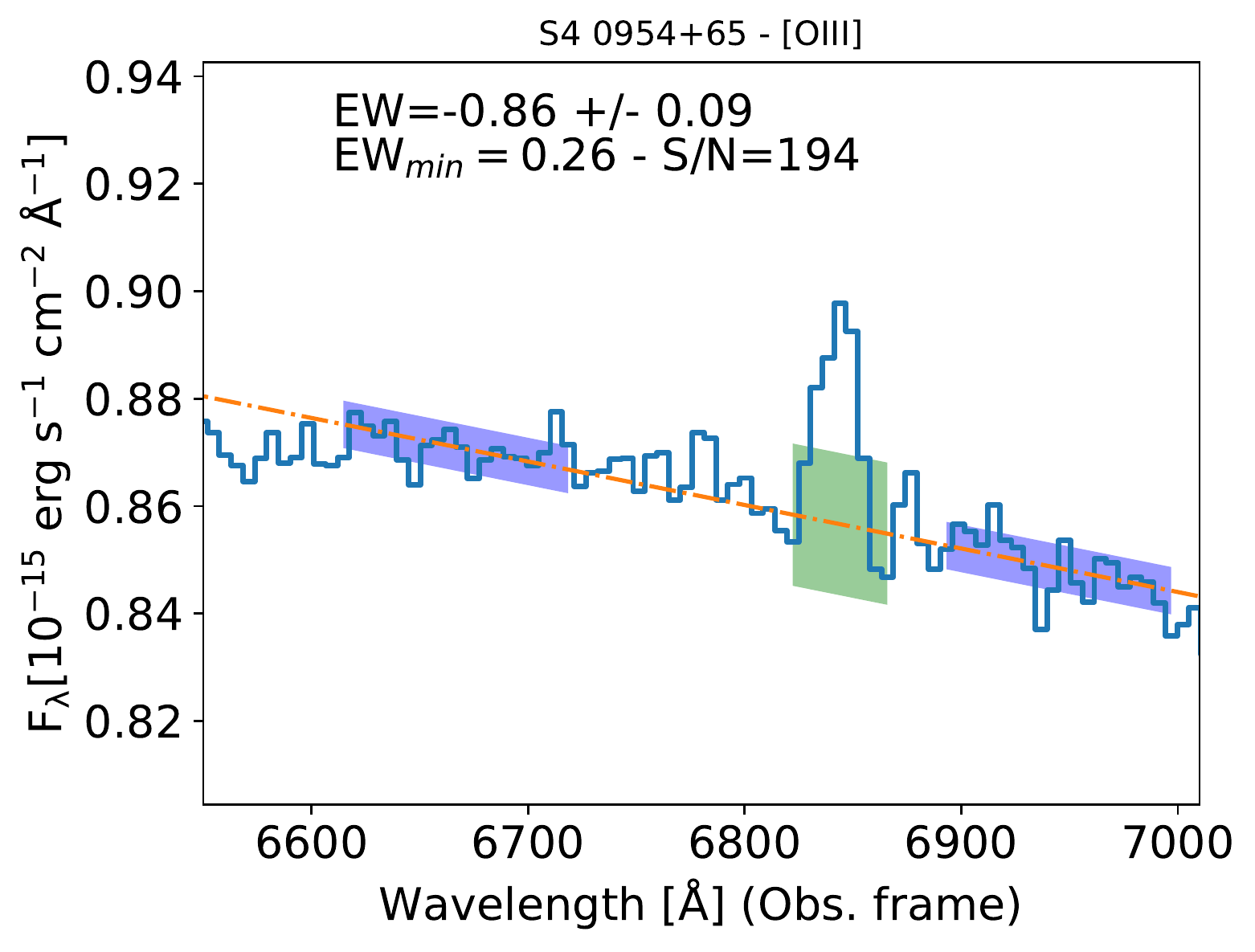}
\includegraphics[scale=0.5]{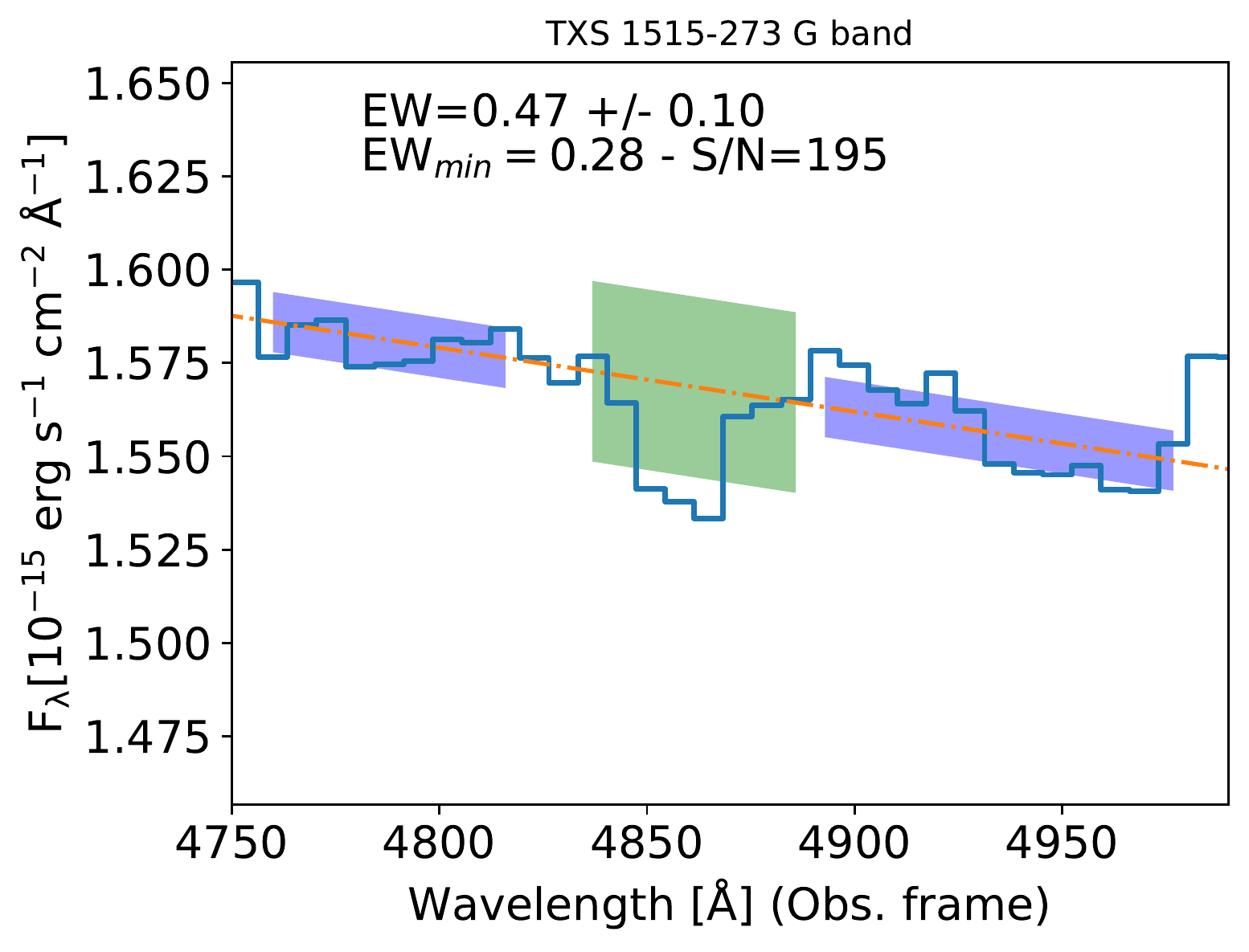}
\includegraphics[scale=0.5]{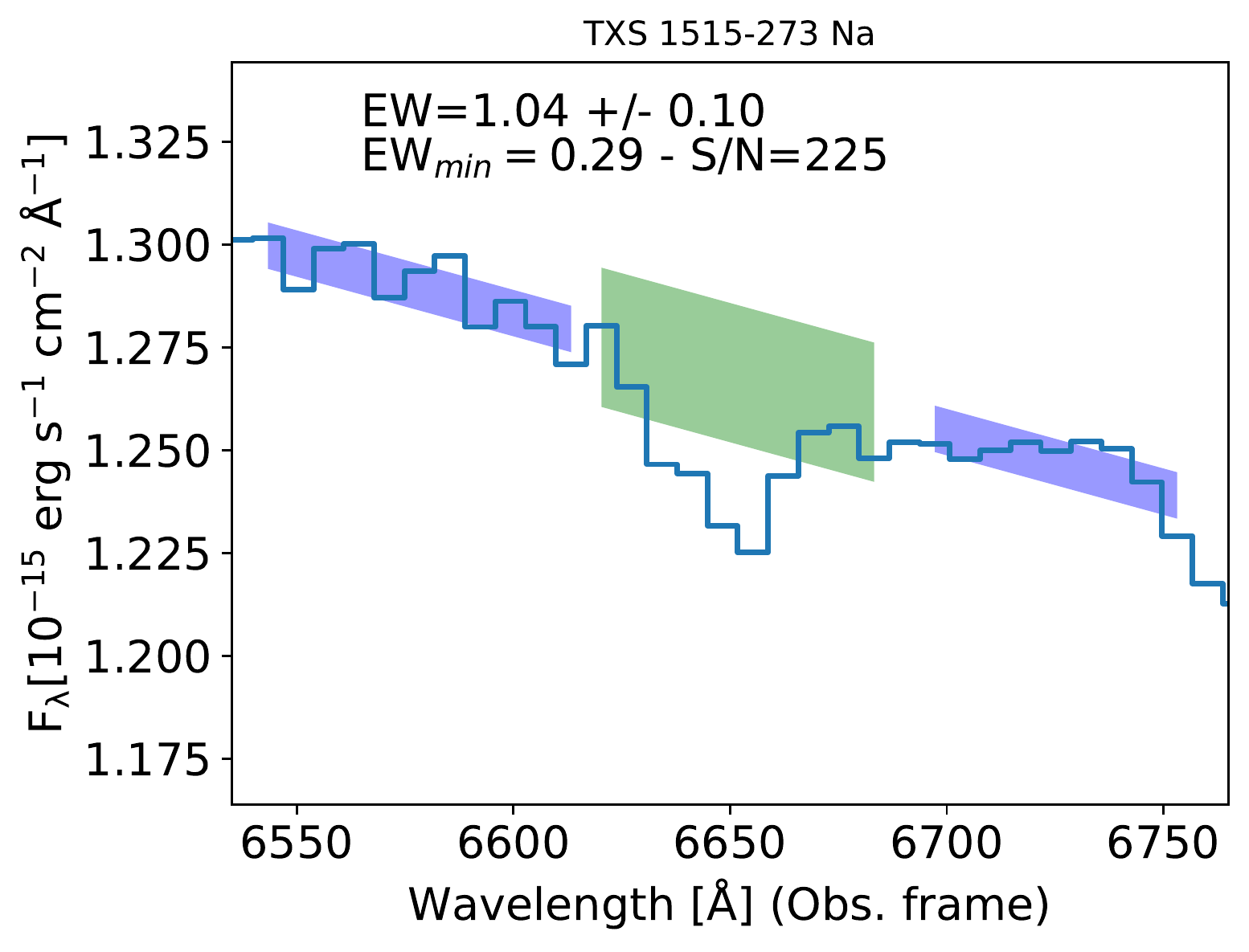}
\caption{Measurement of the EWs of several spectral features in the optical spectra of \s\  (top panels) and \txs\ (bottom panels). The central shadow region represents the region where the EW is computed and the regions at both sides represent the continuum regions. The vertical size of the continuum regions represents the value of the noise (1$\sigma$), whereas at the feature window indicates 3 times the value of the noise (3$\sigma$). The detection of a spectral feature will be more reliable when more flux is contained outside the shaded region. In the case of the [O~III] region, the adjacent blue continuum to the 5007\,\AA\ line is avoided due to the possible contamination of the [O~III] 4959\,\AA\ emission line.  }
\label{fig:EWs_S4_TXS}
\end{figure*}

The method to measure the $\rm{EW}_{min}$ follow in this work measures the EW within the specified wavelength range of a set of spectra obtained by Monte Carlo simulation. We first fit the continuum in the adjacent regions and determine the noise-per-pixel as explained above.  Then the continuum is interpolated within the feature spectral range and a set of Monte Carlo simulations are obtained using as noise for each pixel a random normal distribution. The measured EWs follow a distribution centered around zero and its width represents the EW uncertainty. At the same time we measure the EW of the feature from the observed spectrum, and its uncertainty is estimated by Monte Carlo simulations using the same $\mathit{RMS}_c$ as for the minimum EW.
A graphical illustration of our procedure is presented in Fig. \ref{fig:EWs_S4_TXS} for several lines in the spectra of \s\ and \txs. 

We check that our procedure to estimate the minimum EW is equivalent to that proposed by 
\citet{Sbarufatti05} using simulated pure continuum spectra and also compatible with the value computed using the formulae above.  

\subsection{On the featureless spectrum of \s\  obtained by \citet{2015AJ....150..181L}}
\label{appendix_landoni}

\begin{figure*}
\includegraphics[scale=0.5]{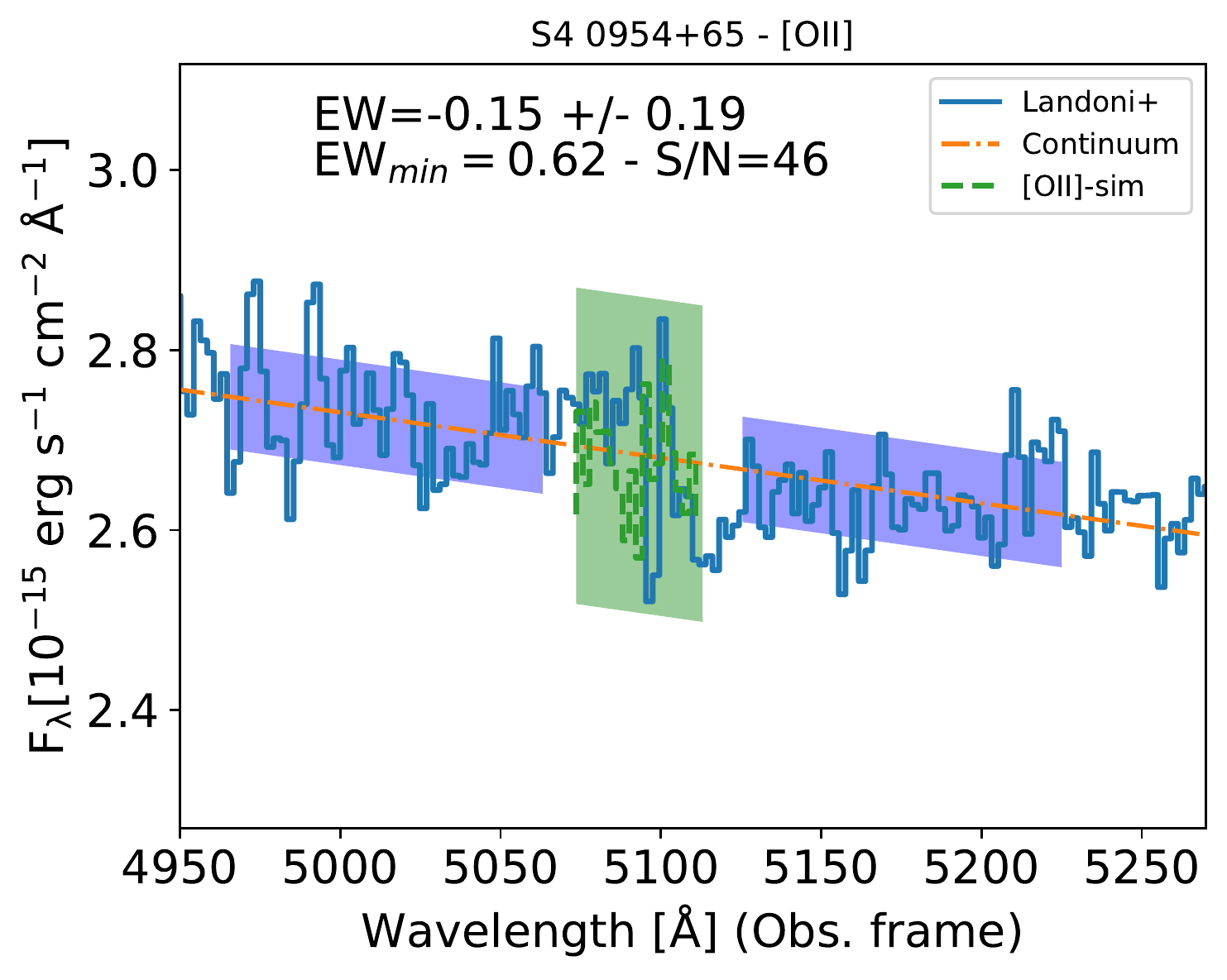}
\includegraphics[scale=0.5]{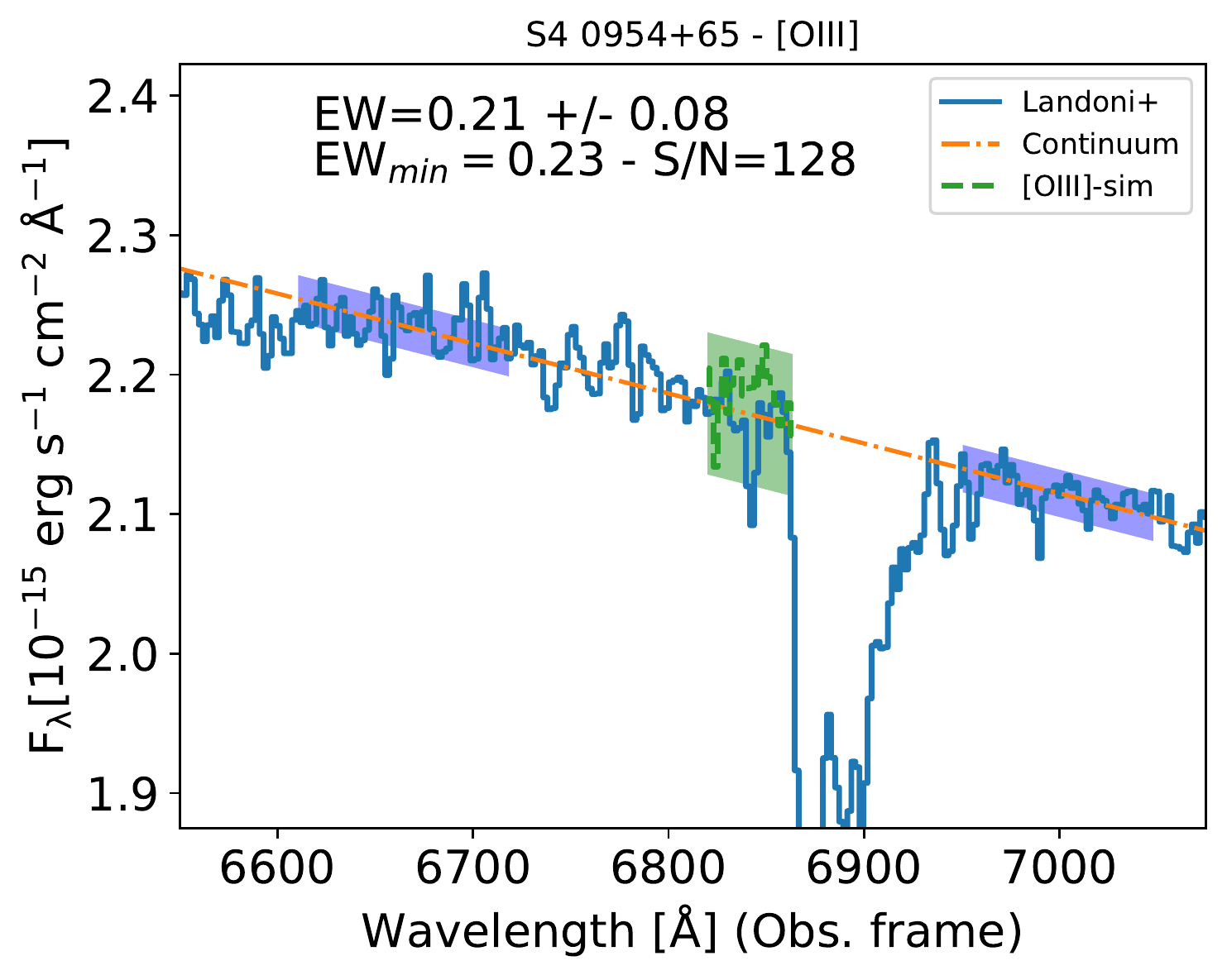}
\caption{Calculation of minimum EW for the spectrum from \citet{2015AJ....150..181L}. The meaning of the shaded regions is the same as in Fig.~\ref{fig:EWs_S4_TXS}. A simulation of the emission lines [O~II]~3726 and [O~III]~5007 is also presented (dashed line), including the noise as measured from the adjacent continuum. The flux of the lines is assumed to be constant and the same as measured in our spectrum corresponding to 2015, May 21st. 
It can be noticed that [O~II] is well below the detection limit, and [O~III] is at the limit of the detection (3$\sigma$) after accounting for the flux difference between the two epochs.}
\label{fig:EWs_Landoni15}
\end{figure*}

The redshift determination of \s\ has been a controversial topic, despite some early claims by \citet{1993A&AS...98..393S,1986AJ.....91..494L}. 
More recently, \citet{2015AJ....150..181L} obtained a high S/N ratio spectrum of \s\ on the night 28-Feb-2015 using the spectrograph OSIRIS at GTC. However, in this spectrum  
the [O~II]~3726\,\AA\ and [O~III]~5007\,\AA\ features are not detected, casting some doubts on previous claims. Here we check that the GTC non-detection is compatible with our results, and is produced by an increased dilution due to an enhancement of the non-thermal emission by almost a factor 3 with respect to the observation in May 2015 (reported in Table~\ref{tab:emission_lines_s4}). Under these assumptions, the expected EWs are -0.13\,\AA\ and -0.29\,\AA\ for the emission lines [O~II] and [O~III], respectively. 
The GTC spectrum was retrieved from the ZBLLAC database\footnote{https://web.oapd.inaf.it/zbllac/}. We re-scale it in flux using the photometry as reported by \citet{2015AJ....150..181L} to have the total source flux. In addition, we apply the correction for interstellar reddening. Using this re-calibrated spectrum we compute the minimum EW, resulting in the values 0.63\,\AA\ and 0.21\,\AA\ around [O~II] and [O~III], respectively (see Fig. \ref{fig:EWs_Landoni15}). Thus the minimum EW around [O~II] is larger than the line expected value (-0.13\,\AA) and the one around [O~III] is of the same order (-0.29\,\AA). Fig. \ref{fig:EWs_Landoni15} also includes a simulation of the expected emission line feature assuming the same Gaussian parameters as found in our observations, but considering the noise of the OSIRIS spectrum. It can be seen that for both cases the scaled spectral lines are not above the $3\sigma$ CL for detection of the line. We also remark that the claim by \citet{2015AJ....150..181L} for $z \geq 0.45$ is based on stellar absorption lines, which are expected at a fainter level than the observed emission lines.


\bsp	
\label{lastpage}
\end{document}